\documentclass[11pt,a4paper]{article}

\usepackage[margin=1in]{geometry}
\usepackage[T1]{fontenc}
\usepackage[utf8]{inputenc}
\usepackage{lmodern}
\usepackage{amsmath,amssymb}
\usepackage{enumitem}
\usepackage{booktabs}
\usepackage{array}
\usepackage{xcolor}
\usepackage{hyperref}
\usepackage{microtype}
\usepackage{csquotes}
\hypersetup{
    colorlinks=true,
    linkcolor=blue!60!black,
    urlcolor=blue!60!black,
    citecolor=blue!60!black
}

\setlist[enumerate]{leftmargin=*, itemsep=7pt, topsep=4pt}
\setlist[itemize]{leftmargin=*, itemsep=6pt, topsep=4pt}

\title{\Huge{From High Energy Astrophysical Neutrinos to the Nobel}\\
\Large{\textit{The History, Legacy and Science Behind the 2026 Nobel Prize in Physics}}}
\author{Dr. Maibam Ricky Devi\\
\\
\textit{Department of Physics, Gauhati University}\\
\textit{Guwahati-781014, Assam, India}\\
\textit{Email: deviricky@gmail.com}}
\date{}
\begin{document}

\maketitle
\section*{The Ghost Particle}

This scientific voyage begins in 1930 with a class of elusive, chargeless elementary particles with infinitesimally small masses called \enquote{neutrinos}. Neutrinos travel through the cosmos without being bent by magnetic fields or absorbed or scattered by intergalactic matter and are hence often called \enquote{ghost particles}. In fact, billions of neutrinos pass through our bodies unscathed every second. This feebly interacting nature of neutrinos with matter can make them notoriously hard to detect. However, the reticence of neutrinos that makes them almost impossible to detect also makes them invaluable astronomical messengers, as they can carry information from the deep core of our galaxy to obscure supermassive black holes without any significant deflection. Originally proposed by Wolfgang Pauli to account for the missing energy in beta decay, neutrinos now play an exceptionally important role in deciphering some of the most exciting conundrums of the universe. This makes neutrinos some of the most unfathomable elementary particles, studied extensively in both astrophysics and particle physics.

\section*{The Big Idea: Neutrino Telescope}

Unlike traditional telescopes, neutrino telescopes are not designed to look up into the sky to detect light; rather, they look for neutrinos below the Earth's surface. However, building a neutrino telescope can be quite a daunting task. It employs a kilometer-scale detector buried more than a mile beneath the surface, which uses a cubic kilometer of diamond-clear ice or ocean water as the interaction target to detect ultra-high-energy neutrinos in the PeV energy range coming from distant astronomical objects \cite{Ahlers:2018fkn}. The naturally dense ice or water essentially makes the background free of high-energy events, allowing the neutrino telescope to detect cosmic neutrinos from all possible zenith angles. The IceCube Observatory is one such neutrino telescope, but definitely not the first of its kind.

\section*{The Physics: When the Ghost Escapes with a Flash}

When a muon neutrino interacts with a nucleon in ice or around the detector, it undergoes inverse beta decay to produce a muon (a charged particle), which travels in nearly the same direction as its parent particle. As it does so, it produces a cone of pale blue light known as Cherenkov radiation, named after the Russian physicist Pavel Alekseyevich Cherenkov, who shared the 1958 Nobel Prize in Physics for its discovery. This form of radiation occurs when a charged particle travels faster than the speed of light in a refractive medium. Here, it does not violate Einstein's special theory of relativity (ESTR), because ESTR was formulated in the context of a vacuum. Cherenkov radiation is analogous to the sonic boom produced by jets: instead of breaking the sound barrier, the charged particle produces a cone of eerie blue light due to the short wavelengths associated with its high-frequency radiation. The digital optical modules (DOMs) installed in IceCube detect the blue flashes, and the direction and energy of the original neutrino can be reconstructed by monitoring the way the particle passes through the grid of sensors \cite{Deppisch2019}. This makes IceCube a neutrino telescope.

\section*{It All Started with AMANDA}

Before IceCube came into existence, there was AMANDA—the Antarctic Muon And Neutrino Detector Array \cite{bowen2017telescope}, \cite{Ereditato2018}. The AMANDA Collaboration, the experimental predecessor of IceCube, officially began in 1992, having been envisaged by the DUMAND (Deep Underwater Muon and Neutrino Detector) Collaboration. The collaboration was led by Bob Morse, the principal investigator (PI) and experimentalist, and Francis Halzen, the co-PI and theorist, both hailing from the University of Wisconsin. The project was funded by the National Science Foundation, while the drilling was paid for by PICO (Polar Ice Coring Office).

In the Antarctic season of 1993/1994, the first physical prototype array, AMANDA-A, was deployed by drilling 800--1000 m-deep holes with pressurized hot water in the 3-km-thick ice layer located a few hundred meters away from the Amundsen--Scott South Pole Station. This prototype array comprised 80 OMs on 4 strings. AMANDA was the test bed for the technology that would monitor the East Antarctic ice, weighing around 6 million tons. The master driller who made this technical challenge possible was Bruce Koci. Although Bruce Koci did an excellent job drilling and deploying the hardware, the team \enquote{goofed up} by deploying it in shallow Antarctic ice that contained many air bubbles. The optical scattering from these bubbles led to significant delays in photon propagation, potentially worsening the angular resolution of the detectors (2--2.5$^\circ$), with the original directional information of the light being essentially lost.

In the spring of 1988, a private collaboration meeting was held between the founding institutions (1990)—Wisconsin, Cal Berkeley, and Irvine—and the other collaborating groups from Stockholm and Uppsala Universities (which joined in 1992), as well as a high-energy physics institute from the former East Germany. The purpose of the meeting was to extend AMANDA to kilometer-scale dimensions, directing it toward a neutrino telescope with better sensitivity and higher angular resolving power. This would provide a higher probability of neutrino detection, as neutrinos would have a greater chance of interacting within a larger volume of ice. Thus, it would not be wrong to say that AMANDA was the heart and soul of the IceCube Neutrino Telescope. AMANDA's final point-source analysis included 6,595 neutrinos detected between 2000 and 2006, before the detector was decommissioned in 2009.

\section*{Building the Icy Dark Room}

In October 2000, IceCube was approved under the Clinton Administration, followed by Francis Halzen becoming the PI of this mega-project in 2001. In 2002, it started receiving funds from the NSF. IceCube was the successor and extension of AMANDA, buried more than a mile beneath the Antarctic ice. In 2004/2005, strings carrying DOMs (Digital Optical Modules) were deployed into holes at a depth of 2.5 km, drilled in a hexagonal grid pattern using a hot-water drill. IceCube had a total of 86 strings equipped with 5,160 DOMs in the ice and 320 DOMs installed in IceTop, along with thousands of optical sensors, by the end of its construction. IceCube was once called the \enquote{weirdest} of the seven wonders of modern astronomy by Scientific American. Building IceCube was a huge engineering challenge, as the collaborators couldn't work beneath the thick ice sheets while building it. Francis Halzen, the Belgian-born theoretical physicist at the University of Wisconsin who envisioned the idea of IceCube, aptly described IceCube as being like a telescope built inside a darkroom.

IceCube can detect high-energy (peta-electron-volt) muons by either the traditional approach, where long muon tracks are detected from neutrino interactions outside the detector, or by High-Energy Starting Events (HESE), which identifies neutrinos from all directions that interact inside the detector. The charged muons emitting Cherenkov radiation can travel several hundred meters before being detected by DOMs. By simply measuring the intensity and arrival time of the light, the direction and energy of the charged particles can be reconstructed. On January 18, 2011, John Jacobsen, a computer wizard and student of Francis Halzen, broadcast an e-mail to the collaborators under the title “IC79 est morte, vive IC86,” announcing the first flight of the IceCube detector. In April 2011, Francis Halzen went out on a limb before the audience at the IceCube commissioning gala in Madison and predicted that IceCube would have a five-sigma discovery within five years.

\section*{The Rise of New Astronomy: “Bert, Ernie and the Billion-Electron-Volt Surprise}

Francis Halzen’s prediction became a reality in 2011–2012, when the IceCube Neutrino Observatory detected two extraordinarily energetic neutrino events, \enquote{Bert} and \enquote{Ernie}, named after the Sesame Street duo \enquote{Bert and Ernie}. In August 2011, Bert was detected at the center of the IceCube array and represented, at the time, the highest-energy neutrino ever observed. A few months later, on January 3, 2012, Ernie was detected with even higher energy than Bert.\\
\\
 In May 2012, IceCube collaborator Aya Ishihara identified these two remarkable events in the data. Subsequently, in October 2012, IceCube collaborators Nathan Whitehorn and Claudio Kopper identified 26 additional events with energies comparable to those of Bert and Ernie. This brought the total number of high-energy events to 28. In May 2013, at a symposium in Madison, Wisconsin, the IceCube Collaboration presented these 28 events to the scientific community, describing them as evidence for the first detection of high-energy neutrinos originating beyond the Solar System. The discovery marked a major milestone in the development of a new branch of astronomy, \enquote{The Neutrino Astronomy}. 
 \\
 In November 2013, the collaboration published its findings in Science, widely regarded as marking the beginning of a new era of neutrino astronomy. The significance of the discovery was recognized internationally. In December 2013, Physics World (UK)  named IceCube’s discovery of cosmic neutrinos its \enquote{Breakthrough of the Year}. In September 2015, Francis Halzen was awarded the \enquote{Balzan Prize}, further recognizing his pioneering contributions to neutrino astronomy and the development of the IceCube Observatory.

\section*{The Scientific Payoff: Neutrinos to Nobel}

The strong conviction by Prof. Francis Halzen to transform a huge block of Antarctic ice into the largest and the most successful neutrino detector in the world and to work on it with relentless grit and determination to make it come true since Day 1 of the AMANDA project, led him to win the \textbf{2026 Nobel Prize in Physics}. This is the story of some great scientists and engineers who dared to dream the impossible and made it happen through bold decisions, bold ideas, and innovative engineering heralding a new branch of astronomy. This leaves an enduring legacy of inspiration for future generations of aspiring scientists, who will look up to these scientific pioneers as role models and strive to follow in their footsteps. 
\bibliographystyle{JHEP_improved}
\bibliography{ref.bib}

\end{document}